# Magnesium-intercalated graphene on SiC: highly n-doped air-stable bilayer graphene at extreme displacement fields

*Antonija Grubišić-Čabo[1*†], Jimmy C. Kotsakidis[1], Yuefeng Yin[2,3], Anton Tadich[3,4,5], Matthew Haldon[1], Sean Solari[1], Iolanda di Bernardo[1,3], Kevin M. Daniels[6], John Riley[5], Eric Huwald[5], Mark T. Edmonds[1,3], Rachael Myers-Ward[7], Nikhil V. Medhekar[2,3], D. Kurt Gaskill[7], Michael S. Fuhrer[1,3*]*

1. School of Physics and Astronomy, Monash University, Clayton VIC 3800, Australia

2. Department of Materials Science and Engineering, Monash University, Clayton VIC 3800, Australia

3. Centre for Future Low Energy Electronics Technologies, Monash University, Clayton VIC 3800, Australia

4. Australian Synchrotron, Clayton VIC 3168, Australia

5. Department of Physics, La Trobe University, Bundoora VIC 3086, Australia

6. Department of ECE, University of Maryland, College Park, Maryland 20742, USA

7. U.S. Naval Research Laboratory, Washington D.C. 20375, USA

We use angle-resolved photoemission spectroscopy to investigate the electronic structure of bilayer graphene at high n-doping and extreme displacement fields, created by intercalating epitaxial monolayer graphene on silicon carbide with magnesium to form quasi-freestanding bilayer graphene on magnesium-terminated silicon carbide. Angle-resolved photoemission spectroscopy reveals that upon magnesium intercalation, the single massless Dirac band of epitaxial monolayer graphene is transformed into the characteristic massive double-band Dirac spectrum of quasi-freestanding bilayer graphene. Analysis of the spectrum using a simple tight binding model indicates that magnesium intercalation results in an n-type doping of $2.1 \times 10^{14}$ cm$^{-2}$, creates an extremely high displacement field of 2.6 V/nm, opening a considerable gap of 0.36 eV at the Dirac point. This is further confirmed by density-functional theory calculations for quasi-freestanding bilayer graphene on magnesium-terminated silicon carbide, which show a similar doping level, displacement field and bandgap. Finally, magnesium-intercalated samples are surprisingly robust to ambient conditions; no significant changes in the electronic structure are observed after 30 minutes exposure in air.

INTRODUCTION

Graphene, a single layer of $sp^2$ bonded carbon atoms [1], has an exceptionally high intrinsic electrical conductivity [2], yet is nearly 98% transparent to light [3] across a broad spectrum of wavelengths, making it attractive as a transparent conductor for a variety of applications.

Moreover, doping may be used to significantly modify graphene's electrical and optical properties. Graphene's conductivity can be tuned dramatically with doping [1], and in highly doped graphene achieved via chemical means [4,5] conductivity can often reach values near the intrinsic limit set by room temperature acoustic phonon scattering. Doping can be used to alter graphene's workfunction [6], which can be exploited to make new types of electronic devices [7] or more efficient contacts to semiconductors [8]. Doping also alters graphene's optical absorption properties. For example, Pauli blocking, where interband optical transitions for energies lower than twice the Fermi energy are forbidden, causes an increase in transparency, an effect which can be exploited for optoelectronic switching [9] or increased performance in transparent conductors [5,10]. In bilayer graphene, doping can produce a displacement field which opens a bandgap at the Dirac point, additionally altering the electronic and optical properties [11–17].

A variety of approaches have been used to tune graphene's properties via doping, including field-effect gating [6,18–21], electric double layer gating [15,22–25], electrolytic gating [26–29], chemical substitution [30–34], adsorption [11,35–41], and intercalation [42–44,51–55]. Among these, chemical doping offers a simple, powerful approach to create highly-doped graphene layers which can be incorporated as transparent conductors, electrodes or optical elements in a wide variety of device structures. To be widely applicable, the chemical doping approach should result in a highly-doped graphene layer which is stable under processing conditions such as ambient exposure and high temperature. Several chemical doping approaches have been demonstrated to successfully produce stable highly p-doped graphene [50,56–59] with p-type carrier density exceeding $10^{14}$ cm$^{-2}$. Stable n-doped graphene is also desirable, particularly for applications requiring low work function (as compared to an increased work function in the case of p-doping). However the production of stable n-doped graphene has been more difficult, with only a few demonstrations [44,60,61]. In large part, difficulty in producing stable n-doped graphene is due to the highly reactive and air-unstable nature of n-type dopants. Despite this, highly air-stable, n-doped single-layer graphene was obtained by CsCO$_3$ [60], ZnO doping [61], attaining electron concentrations of $2.2 \times 10^{13}$ cm$^{-2}$ and $>5.76 \times 10^{12}$ cm$^{-2}$, respectively. These values do not significantly exceed the natural doping found in epitaxial monolayer graphene (EMLG) on silicon carbide [62], so achieving extremely high and stable n-doped graphene remains an open challenge.

Here, we use angle-resolved photoemission spectroscopy (ARPES) to study the recently reported quasi-freestanding bilayer graphene on a magnesium-terminated SiC substrate (Mg-QFSBLG) created by magnesium intercalation of epitaxial monolayer graphene on 6H-silicon carbide (SiC) [63]. Analysis of the electronic spectrum using a simple tight binding model indicates high n-doping ($>2 \times 10^{14}$ cm$^{-2}$). The exceptionally high displacement field produced by the charge transfer from the intercalated magnesium to graphene opens a large (0.36 eV) bandgap at the Dirac point. Moreover, the high level of n-doping is stable after heating to 350 ºC, as well as 30 minutes of exposure to air. The electronic spectrum of the highly n-doped bilayer graphene is well described by a simple tight-binding model for bilayer graphene with displacement field. First principles density-functional theory (DFT) calculations corroborate that magnesium intercalation produces quasi-freestanding bilayer graphene with good agreement in doping level, Fermi energy, and bandgap to our experimental values.

METHODS

Epitaxial monolayer graphene samples of nominally sub-monolayer coverage were grown on a silicon face of a semi-insulating 6H-SiC substrate by silicon sublimation from the SiC, as described in Ref. [64]. Sample preparation, ARPES, and low-energy electron diffraction (LEED) measurements were carried out at the Toroidal Analyzer endstation at the Soft X-ray Beamline of the Australian Synchrotron. Samples were introduced to ultra-high vacuum (UHV, base pressure of $1 \times 10^{-10}$ mbar), and annealed over night at 500 °C. Sample cleanliness was confirmed by LEED and ARPES. A magnesium effusion cell was baked at 150 °C overnight and outgassed at 415 °C. Once the pressure reached $1 \times 10^{-7}$ mbar, the effusion cell was inserted into the UHV preparation chamber. Magnesium (1/8 inch turnings, 99.95%, Sigma Aldrich) was intercalated following the recipe from Ref. [63]: Magnesium was evaporated for 25 min, with the magnesium cell held at 400 °C, and deposited on the graphene/SiC substrate held at room temperature in thickness of 188 Å, as determined by quartz crystal microbalance. Following the deposition, the graphene/SiC substrate was annealed at 350 °C for 30 min to facilitate magnesium intercalation under the graphene. For the air exposure experiment, after 30 minutes of air exposure, the sample was reintroduced to UHV and annealed at 350 °C for several hours prior to measurements.

Structural characterisation of the samples was undertaken using LEED (OCI$^{TM}$ 3 grid reverse view optics, 200 μm spot size) at room temperature, at energies between 56 eV and 200 eV, *in-situ* in the endstation used for ARPES. ARPES measurements used a toroidal-type angle-resolving endstation [65] at the Soft X-Ray Beamline of the Australian Synchrotron. All ARPES data was taken at room temperature and a photon energy ($hv$) of 100 eV using linearly polarised light at normal incidence to the sample, with a beam spot size of 100 μm × 60 μm. The binding energy ($E_{Bin}$) scale for all spectra are referenced to the Fermi energy ($E_F$), determined using the Fermi edge of an Au foil reference sample in electrical contact with the sample. The toroidal analyser permits all polar ($\theta$) emission angles (-90° to +90°) to be measured along a high-symmetry azimuth ($\varphi$) of the surface containing the $\bar{\Gamma}$ point. The unique geometry therefore allows for measurement of the Dirac cone along the $\bar{K} - \bar{\Gamma} - \bar{K}$ direction without the need for complex alignment of the spectrometer. A simple rotation of the sample in azimuth was then used to measure the Dirac point along the direction perpendicular to the $\bar{K} - \bar{\Gamma} - \bar{K}$ direction. Using this latter method avoids the well-known intensity suppression of half of the Dirac cone seen when measuring along the $\bar{K} - \bar{\Gamma} - \bar{K}$ using this polarization geometry [66] and provides a more robust means of determining the Dirac point and carrier velocities. The measurement direction perpendicular to the $\bar{K} - \bar{\Gamma} - \bar{K}$ direction; however, exhibits a lower $k_\parallel$ instrumental resolution than in the $\bar{K} - \bar{\Gamma} - \bar{K}$ direction, resulting in higher than normal momentum broadening in the data. The effect is due to the finite-size analyser slit that is used when measuring along the bandstructure along the azimuthal direction. The result is approximately an order of magnitude decrease in the instrumental angular resolution compared to scanning $k_\parallel$ using the polar emission angle [65]. The contribution to the momentum uncertainty due to the angular resolution along $\varphi$ is estimated to be ~ 0.1 Å$^{-1}$, compared with ~ 0.01 Å$^{-1}$ for measurements taken along $\theta$. In both measurement directions, however, the energy resolution is ~ 100 meV. Data taken along the $\overline{K} - \overline{\Gamma} - \overline{K}$ direction can be found in the Supplementary Material, section 2.

First principles density-functional theory calculations were implemented using the Vienna ab initio Simulation Package (VASP) to calculate the electronic structure of Mg-QFSBLG [67]. The Perdew-Burke-Ernzehof (PBE) form of the generalized gradient approximation (GGA) was used to describe electron exchange and correlation [68]. A semi-empirical functional (DFT-D2) was employed to describe van der Waals interactions in the system [69]. The kinetic energy cut-off for the plane-wave basis set was set to 500 eV. We used a 9 × 9 × 1 Γ-centered k-point mesh for sampling the Brillouin zone. The unfolded band structure and Fermi surface were obtained using the KPROJ program based on the *k*-projection method [70,71].

RESULTS AND DISCUSSION

Intercalation is a method commonly used to tailor the properties of graphene [42–45, 49–53,72–76]. The advent of epitaxial graphene on SiC has offered new opportunities for intercalation, as many species which will not intercalate graphite [77,78] will in fact intercalate the graphene-SiC interface [43,44,51,79,80] and alter the properties of the graphene overlayer(s). Magnesium is one such species which does not intercalate graphite [81], and therefore is not expected to intercalate in the galleries between graphene layers, however was recently observed to intercalate EMLG on SiC [63]. In this case, intercalation is possible due to the different chemical nature of the silicon-graphene interface, where the silicon is bonded to the first carbon layer, known as the buffer layer (Figure 1a). During the intercalation process, magnesium, rather than intercalating the graphene layer, goes under the buffer layer and sits on top of SiC [63], as shown in Figure 1b. Once magnesium is intercalated under the buffer layer, it effectively cuts the bonds between the carbon atoms in the buffer layer and silicon dangling bonds, thus turning the buffer layer into a free graphene layer, and by extension transforming EMLG into Mg-QFSBLG.

In order to confirm that EMLG is converted structurally to Mg-QFSBLG we use LEED. Figures 1c and 1d show LEED images before and after magnesium intercalation, respectively. Before intercalation, we observe the characteristic LEED pattern of EMLG, with $(6\sqrt{3} \times 6\sqrt{3})R30°$ reconstruction relative to the SiC lattice characteristic of the buffer layer (orange circles, Figure 1c) in addition to the $(1 \times 1)$ graphene and $(1 \times 1)$ SiC spots (green and gray circle, respectively, Figure 1c). After intercalation, the $(6\sqrt{3} \times 6\sqrt{3})R30°$ spots are greatly reduced in intensity, and graphene $(1 \times 1)$ spots are significantly more pronounced than $(1 \times 1)$ SiC spots, indicating a reduced interaction with the substrate [62]. Additional $(\sqrt{3} \times \sqrt{3})R30°$ spots with respect to the $(1 \times 1)$ SiC spots are visible (yellow circle, Figure 1d) after intercalation and are attributed to the formation of the magnesium silicide-like surface reconstruction under the graphene [63]. The first principles calculations support the interpretation that EMLG is converted to the Mg-QFSBLG heterostructure, shown in Figure 1b, with the energy of Mg-intercalated structure lower than the energy of crystalline Mg on epitaxial monolayer graphene by 1.18 eV. Additional LEED data and more details on the calculation of relative energies can be found in the Supplementary Material, section 1 and 5, respectively.

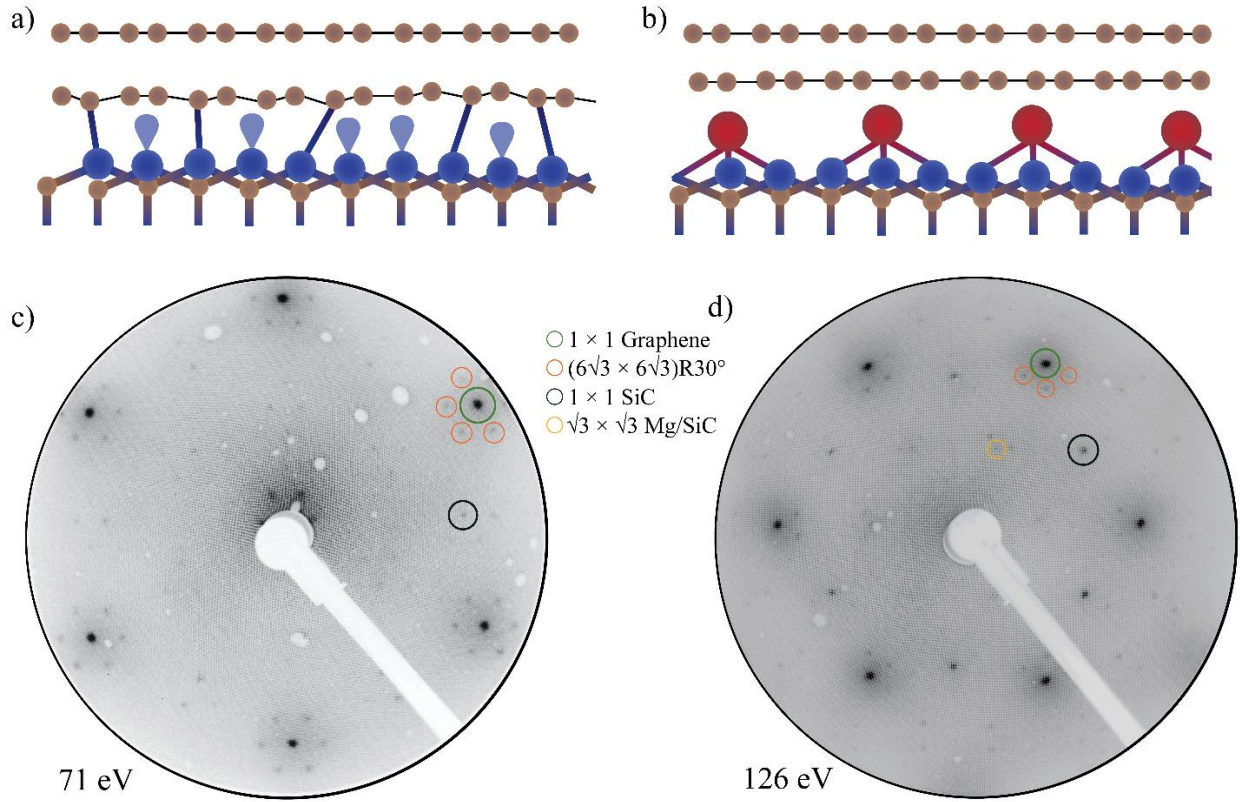

**Figure 1.** Magnesium intercalated epitaxial monolayer graphene. Sketch of a) epitaxial monolayer graphene on SiC and b) magnesium-intercalated quasi-freestanding bilayer graphene on SiC. Brown spheres: carbon; blue spheres: silicon; blue lobes: silicon dangling bonds; red spheres: magnesium. LEED image of epitaxial monolayer c) before and d) after magnesium intercalation. LEED images taken at 71 eV and 126 eV, respectively, on the same sample. Sample was remounted between the LEED measurements. Green circle: (1 × 1) graphene lattice; gray circle: (1 × 1) SiC lattice; orange circles: (6√3 × 6√3)R30° reconstruction relative to SiC arising from the buffer layer, yellow circle: (√3 × √3) R30° reconstruction of SiC surface by magnesium.

LEED itself, being a structural technique, cannot provide insight into the effect of magnesium intercalation on the electronic structure of graphene. To assess electronic structure changes, a more direct probe of the electronic structure is needed. One such probe is the ARPES technique, which can directly visualise the electronic structure of materials and give information about doping, bandgap, number of layers, and many-body interactions [82,83].

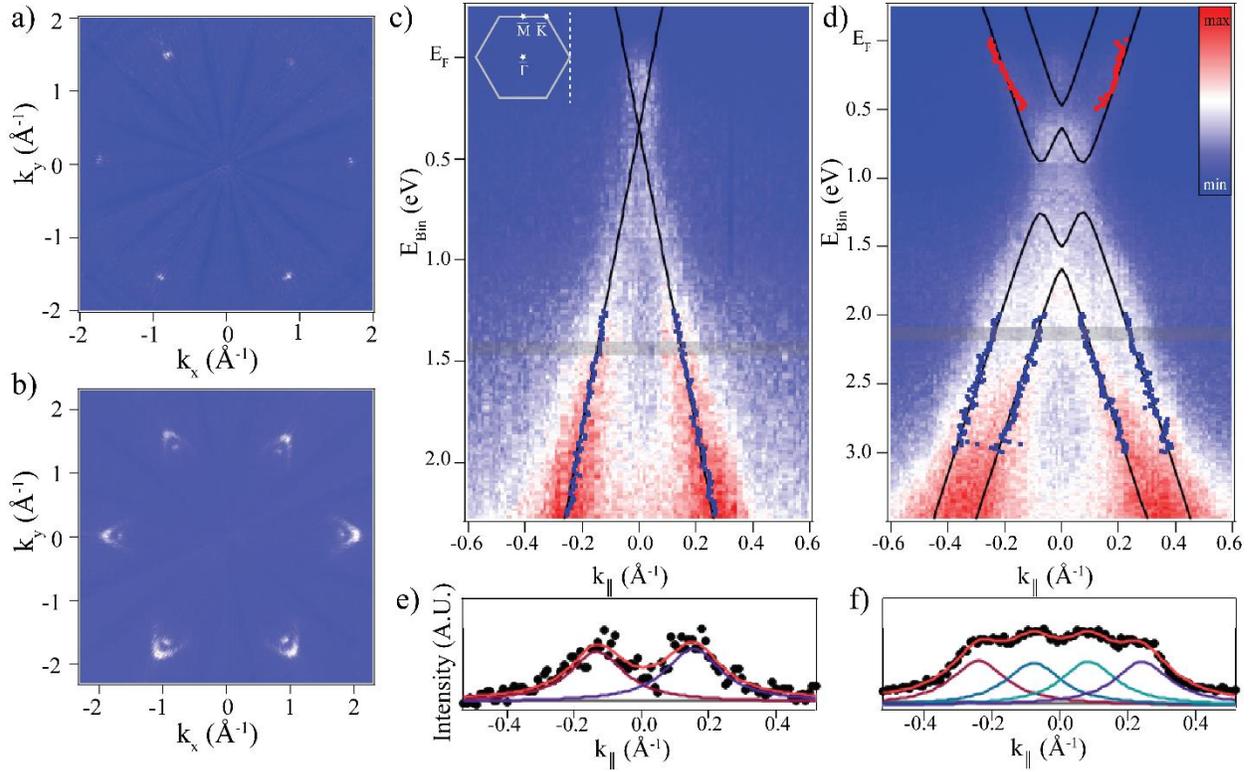

**Figure 2.** Electronic structure of graphene on SiC before and after magnesium intercalation. Constant energy surfaces taken at the Fermi level showing a) monolayer graphene Fermi surface and b) bilayer graphene Fermi surface following magnesium intercalation. Band dispersion of c) monolayer graphene before and d) bilayer graphene after magnesium intercalation. Blue and red markers are extracted band positions from momentum distribution curves (MDCs). Overlaid in black is c) linear fit and d) tight-binding model for $U = 0.87$ V, defined in the Eq.1 of the main text. e) and f) are extracted MDCs from the grey shaded area in c) and d), respectively, showing

two (four) bands as expected for monolayer (bilayer) graphene, Bands were averaged in a 50 meV window, taken 1.1 eV below extrapolated Dirac point. All data taken at $h\nu = 100$ eV and at room temperature.

Figure 2 shows ARPES measurements, before and after magnesium intercalation. Figures 2a and 2b, show the Fermi surface (spectral weight as a function of in-plane momentum at constant energy at the Fermi level) of clean (non-intercalated) EMLG and Mg-QFSBLG, respectively. Here, the unique toroidal analyser geometry [65] enables the detection of a full hemisphere (i.e. a 180° photoelectron emission window), which samples a wide k-space. The differences in the Fermi surfaces are easily seen: Prior to the intercalation (Figure 2a) the Fermi surface consists of an individual circular feature characteristic of the single Dirac cone of EMLG [84]. Following magnesium intercalation (Figure 2b), an additional feature develops and two well-separated Fermi surfaces are clearly visible as a smaller circular feature enveloped by a larger triangular one. This is consistent with the bilayer graphene structure [11]. Note that the absence of the intensity on one side of the Fermi surface contour in the bilayer (and monolayer) graphene case is due to the interference effect from the two atoms in a graphene unit cell [66,84,85].

Figures 2c and 2d show the band dispersion measured perpendicular to the $\bar{K} - \bar{\Gamma} - \bar{K}$ high-symmetry direction, as indicated schematically in the inset of Figure 2c. This direction is chosen because there are no changes in the graphene band intensity along this vector due to matrix-element effects. Before intercalation, Figure 2c, a single set of linearly dispersing bands is visible, as expected for EMLG. The Dirac point position and Fermi velocity of $v_F = (1.17 \pm 0.02) \times 10^6$ m/s, a value similar to typically reported Fermi velocity for EMLG on SiC [86,87], were obtained from the linear fit (black line, Figure 2c) to the band position values

(blue markers) taken from the momentum distribution curves (MDCs). The Dirac point lies below the Fermi level, $E_F - E_D = 0.35 \pm 0.04$ eV, value comparable to typical reported values for EMLG on SiC [35,62,73]. Fermi wavevector is determined to be $0.048 \pm 0.004$ Å$^{-1}$, corresponding to a carrier density of $n = (7.3 \pm 0.6) \times 10^{12}$ cm$^{-2}$.

After intercalation, Figure 2d, two sets of bands are visible, as is expected for bilayer graphene. Red (blue) markers represent the conduction (valence) band position values obtained from the MDCs. These values were fitted to a tight-binding model, Eq. 1, overlaid in black, for bilayer graphene under a perpendicular displacement field, based on Refs. [11,12,88]:

$$\varepsilon_\alpha(k) = \pm \left[ \frac{\gamma_1^2}{2} + \frac{U^2}{2} + \left( v^2 + \frac{v_3^2}{2} \right) k^2 + (-1)^\alpha \sqrt{\Psi} \right]^{1/2}, \quad (1)$$

where α=1,2 is the band index, and

$$\Psi = \frac{1}{4}(\gamma_1^2 - v_3^2 k^2)^2 + v^2 k^2 (\gamma_1^2 + U^2 + v_3^2 k^2) + 2\gamma_1 v_3 v^2 k^3 \cos 3\varphi$$

and $v_3 = \frac{\sqrt{3} a \gamma_3}{2\hbar}$.

Here $k$ is the wavevector, $\varphi$ is the azimuthal angle, $v$ is the band velocity, $U$ is the difference in the onsite Coulomb potential of two graphene layers, $\gamma_1 = 0.4$ eV is the out-of-plane nearest-neighbour interaction parameter, $\gamma_3 = 0.12$ eV is the out-of-plane next-nearest neighbour interaction parameter, $a = 1.42$ Å is the C-C distance in graphene, and $\hbar$ is reduced Planck's constant [11].

From the fit, we obtain the band gap value of $E_G = 0.36 \pm 0.04$ eV, value in agreement with theoretically predicted band gap for bilayer graphene under high displacement field [13], and $v =$

$(0.97 \pm 0.04) \times 10^6$ m/s, same order of magnitude as in Ref. [11]. The tight-binding model includes an interlayer potential difference of $0.87 \pm 0.06$ V, yielding an extremely high displacement field of $2.6 \pm 0.2$ V/nm [89,90]. From the band parameters we obtain the Fermi wavevectors $k_{F,1} = 0.24 \pm 0.01$ Å$^{-1}$ (outer band) and $k_{F,2} = 0.09 \pm 0.01$ Å$^{-1}$ (inner band). We estimate the carrier densities as $n_i = k_{F,i}^2/\pi$ for i = 1,2. Note that the first-order correction to the Fermi wavevector due to trigonal warping is zero along the direction perpendicular to the $\bar{K} - \bar{\Gamma} - \bar{K}$ high-symmetry direction so this provides a good approximation even for the trigonally warped surface. We then find carrier densities $n_1 = (1.83 \pm 0.15) \times 10^{14}$ cm$^{-2}$ (outer band) and $n_2 = (0.26 \pm 0.06) \times 10^{14}$ cm$^{-2}$ (inner band), giving a total carrier density of $n = n_1 + n_2 = (2.1 \pm 0.2) \times 10^{14}$ cm$^{-2}$, and an interlayer difference in carrier density $n_1 - n_2 = (1.6 \pm 0.2) \times 10^{14}$ cm$^{-2}$.

It is expected that the interlayer potential difference responds linearly to the interlayer carrier density difference, i.e. $U = \alpha(n)(n_1 - n_2)$ where the linear response coefficient $\alpha(n)$ depends on the total carrier density $n$ [91]. We observe $\alpha(n = 2.1 \pm 0.2 \times 10^{14}$ cm$^{-2}) = (5.4 \pm 0.9) \times 10^{-12}$ meV·cm$^2$. While $\alpha(n)$ has not to our knowledge been calculated from first principles at the very high carrier densities as in our experiment, Ref. [91] showed that calculation of $\alpha(n)$ using a *GW* approach gives excellent agreement with experiment at low $n$, and furthermore extrapolated their *GW* calculation analytically to high $n$. Using their extrapolation we find $\alpha(n = 2.1 \times 10^{14}$ cm$^{-2}) = 5.9 \times 10^{-12}$ meV·cm$^2$, in excellent agreement with our observation.

The extrapolated Dirac point position for the Mg-QFSBLG, $E_F - E_D$, is $1.07 \pm 0.07$ eV, corresponding to a Fermi level shift of $0.72 \pm 0.08$ eV with respect to the EMLG. Figures 2e and 2f show MDCs obtained from the shaded areas (1.1 eV below the extrapolated Dirac point, bands averaged over a 50 meV binning window) in Figures 2c and 2d, which clearly indicate the presence of two (four) bands, as expected for monolayer (bilayer) graphene. The total carrier density in our system is significantly higher than in the previously reported air-stable n-doped graphene systems [60,61], though higher densities have been achieved in vacuum for example by co-doping graphene by K and Ca [45], or by Cs [42] and Gd doping [43].

The electronic structure of the magnesium-intercalated sample obtained by ARPES measurements can be compared with the first principles DFT calculations for a bilayer graphene system where magnesium atoms are sitting at the interface with the SiC substrate. The heterostructure is modelled using a ($\sqrt{3} \times \sqrt{3}$) SiC supercell and a ($2 \times 2$) graphene supercell with one magnesium atom placed in between the two materials (Figures 3a and 3b). The magnesium atom is located on the C-top location which is found to be stable and the most energetically favorable configuration (see Supplementary Materials and Figure S9 for details). The lattice constant of SiC is unchanged while the graphene is stretched by 7.5%. We calculate the Fermi surface and electronic band structure of the system, Figure 3c and 3d. The calculated Fermi surface (Figure 3c) agrees well with our ARPES spectra, where two features are observed in the Fermi surface: A circular feature belonging to the top graphene layer (red contour lines), and a triangular one coming from the bottom layer (green contour lines). The DFT calculations also reproduce the experimental band dispersion as shown in Figure 3d. The band gap is 0.35 eV, which is in excellent agreement with experimental observations. The doping level obtained from the calculations is $3.6 \times 10^{14}$ cm$^{-2}$, somewhat larger than the experimental value of $2.1 \times 10^{14}$ cm$^{-2}$, while the calculated Fermi energy relative to the Dirac point $E_F - E_D = 0.71$ eV is somewhat smaller than experimental value (1.07 eV). The differences are likely related to the artificial stretching of the graphene lattice by 7.5% which preserves the symmetry of the system but lowers the Fermi velocity by 10% relative to the true value [92].

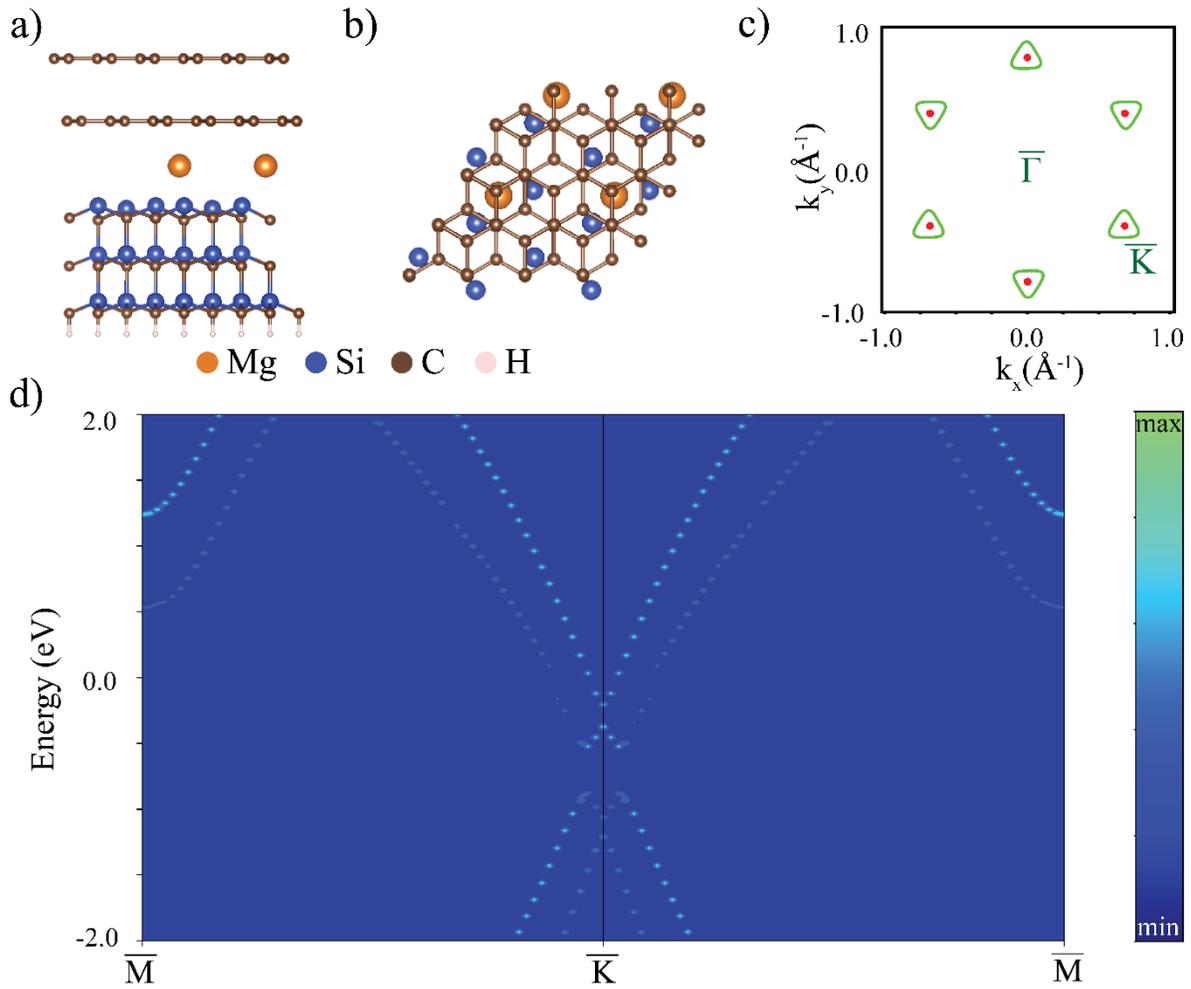

**Figure 3.** First principles DFT calculations of magnesium-intercalated bilayer graphene. Model used in DFT calculations: a) side view and b) top view of the graphene/magnesium/SiC interface. Brown, orange and blue spheres indicate the positions of carbon, magnesium and silicon atoms. Only the topmost silicon atoms of the SiC substrate are shown for clarity. c) Calculated unfolded constant energy slice at the Fermi level of magnesium-intercalated bilayer graphene. Red and green represent contribution from top and bottom graphene layer, respectively. d) Unfolded band dispersion of magnesium-intercalated bilayer graphene.

Highly n-doped graphene/SiC has previously been achieved by depositing or intercalating alkali and alkali-earth metals on graphene [11,43–45,93], however the resulting systems are typically unstable when exposed to air. In our case, magnesium is buried between bilayer graphene and SiC, so it is conceivable that samples could survive air exposure. In order to test air stability, the magnesium-intercalated sample was taken out of UHV and exposed to air for 30 minutes.

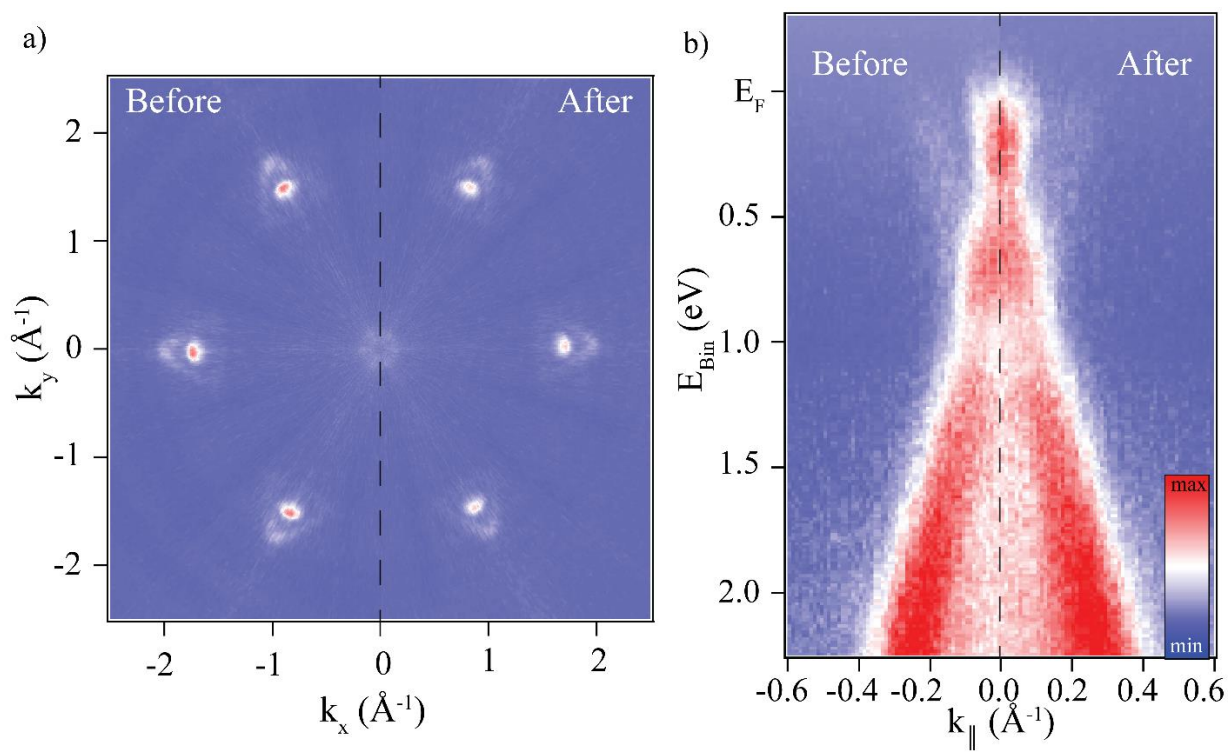

**Figure 4.** Magnesium-intercalated epitaxial monolayer graphene before and after air exposure. Constant energy contours a) and background subtracted band dispersion b) of magnesium-intercalated sample before (left) and after (right) 30 min air exposure. Data taken at $h\nu$=100 eV and at room temperature. Constant energy contour data was averaged around $\bar{K}$ point to increase signal-to-noise ratio.

Figure 4a shows the Fermi surface measured by ARPES before and after 30 minutes air exposure. In both cases, the two clearly separated conduction bands of bilayer graphene are visible, with no significant change in the size of the Fermi surface (directly proportional to doping) after air exposure. Figure 4b shows the electronic dispersion before and after 30 minutes air exposure. No significant changes are observed in the Fermi energy $E_F - E_D$, or the bandgap for our air exposed Mg-QFSBLG, within the experimental resolution. This degree of air stability is surprising for a surface layer and indicates that Mg-QFSBLG created by magnesium intercalation is relatively robust to ambient exposure, which is desirable for designing transparent conducting electrodes [5] with a low-work function. Note that the sample used for the air exposure experiment was a different sample (EMLG with nominally 1 monolayer coverage) than the one for which data is shown in Figure 1 and Figure 2 due to the experimental time constraints. Full LEED and ARPES characterisation of this sample can be found in the Supplementary Material, section 3 and 4, respectively.

CONCLUSIONS

We demonstrate that magnesium intercalation at the interface of SiC and the graphene buffer layer, transforms epitaxial monolayer graphene into quasi-freestanding bilayer graphene, as observed by LEED and ARPES. Once at the interface, magnesium acts as an electron donor and dopes graphene, shifting a Fermi level by 0.72 eV and resulting in an electron carrier density of $n = 2.1 \times 10^{14}$ cm$^{-2}$, proportionate to highest densities achievable with electrolytic gating

($4 \times 10^{14}$ cm$^{-2}$). Magnesium intercalation also creates an extremely high displacement field of 2.6 V/nm, comparable to the largest displacement fields (2.5–3.1 V/nm) obtained in dual gated bilayer graphene FETs. The field of 2.6 V/nm opens a bandgap of 0.36 eV, a value very close to $\gamma_1$ (out-of-plane nearest-neighbour interaction parameter) where the field induced bandgap is expected to saturate, and increases the splitting between the valence (conduction) bands of bilayer graphene. Despite this extremely high displacement field, the electronic structure of the Mg-QFSBLG can still be described with a simple tight-binding model that reproduces well both the bandgap opening and increase in the splitting between bands. First principles DFT calculations are in good agreement and reproduce the experimental band structure well, including the bandgap opening and the increase in the band splitting. An air exposure test shows that the Mg-QFSBLG samples are stable in air for up to 30 minutes, and are thermodynamically stable up to at least 350 °C, suggesting that magnesium-intercalated graphene could be a suitable candidate for application in transparent electrodes and organic opto-electronics.

ASSOCIATED CONTENT

**Supporting Information**.

The following files are available free of charge.

Additional LEED data for sample shown in Figure 1-3; discussion of reduced energy resolution in ARPES data taken along the direction perpendicular to the $\bar{K}$-$\bar{\Gamma}$-$\bar{K}$ direction; ARPES data averaged along the $\bar{K}$-$\bar{\Gamma}$-$\bar{K}$ direction for sample shown in Figure 1-3; additional LEED and ARPES data for sample shown in Figure 4; ARPES data averaged along the $\bar{K}$-$\bar{\Gamma}$-$\bar{K}$ direction for sample shown in Figure 4; details of first principles calculations of the energetics of the Mg

intercalation (PDF)

AUTHOR INFORMATION

**Corresponding Author**

*Antonija@kth.se

* Michael.Fuhrer@monash.edu

**Present Address**

†KTH, Applied Physics, Stockholm SE 114 19, Sweden

**Author Contributions**

The manuscript was written through contributions of all authors. All authors have given approval to the final version of the manuscript.

**Notes**

The authors declare no competing financial interests.

ACKNOWLEDGMENT

This work was supported by the Australian Research Council under awards DP150103837, DP200101345 and FL120100038. This research was undertaken on the Soft X-ray spectroscopy beamline at the Australian Synchrotron, part of ANSTO. JCK gratefully acknowledges support from the Australian Government Research Training Program, and the Monash Centre for Atomically Thin Materials. YY and NM gratefully acknowledge the support from the Australian Research Council (CE17010039) and the computational support from the National Computing Infrastructure and Pawsey Supercomputing Facilities. D.K.G., R.M-W., and K.M.D. acknowledge support by core programs at the U.S. Naval Research Laboratory funded by the Office of Naval Research.

ABBREVIATIONS

EMLG, epitaxial monolayer graphene; ARPES, angle-resolved photoemission spectroscopy; SiC, silicon carbide; Mg-QFSBLG, quasi-freestanding bilayer graphene on magnesium terminated silicon carbide; LEED, low-energy electron diffraction; UHV, ultra-high vacuum; $h\nu$, photon energy; $E_{Bin}$, binding energy; $E_F$, Fermi energy; $\theta$, polar emission angle; $\varphi$, azimuthal angle; VASP, Vienna ab initio Simulation Package; PBE, Perdew-Burke-Ernzhof; GGA, generalized gradient approximation; MDCs, momentum distribution curves.

References

1. Novoselov KS, Geim AK, Morozov S V, Jiang D, Zhang Y, Dubonos S V, et al. Electric Field Effect in Atomically Thin Carbon Films. *Science (80)*. **2004**;306(5696):666–9.

2. Chen J-H, Jang C, Xiao S, Ishigami M, Fuhrer MS. Intrinsic and extrinsic performance limits of graphene devices on SiO2. *Nat Nanotechnol*. **2008**;3(4):206–9.

3. Nair RR, Blake P, Grigorenko AN, Novoselov KS, Booth TJ, Stauber T, et al. Fine Structure Constant Defines Visual Transparency of Graphene. *Science (80)*. **2008**;320(5881):1308.

4. Khrapach I, Withers F, Bointon TH, Polyushkin DK, Barnes WL, Russo S, et al. Novel Highly Conductive and Transparent Graphene-Based Conductors. *Adv Mater*. **2012**;24(21):2844–9.

5. Bao W, Wan J, Han X, Cai X, Zhu H, Kim D, et al. Approaching the limits of transparency and conductivity in graphitic materials through lithium intercalation. *Nat Commun*. **2014**;5(1):4224.

6. Yu Y-J, Zhao Y, Ryu S, Brus LE, Kim KS, Kim P. Tuning the Graphene Work Function by Electric Field Effect. *Nano Lett*. **2009**;9(10):3430–4.

7. Yang H, Heo J, Park S, Song HJ, Seo DH, Byun K-E, et al. Graphene Barristor, a Triode Device with a Gate-Controlled Schottky Barrier. *Science (80)*. **2012**;336(6085):1140–3.

8. Cui X, Lee G-H, Kim YD, Arefe G, Huang PY, Lee C-H, et al. Multi-terminal transport measurements of MoS2 using a van der Waals heterostructure device platform. *Nat Nanotechnol*. **2015**;10(6):534–40.

9. Wang F, Zhang Y, Tian C, Girit C, Zettl A, Crommie M, et al. Gate-Variable Optical Transitions in Graphene. *Science* (80). **2008**;320(5873):206–9.

10. Chen Y, Yue Y-Y, Wang S-R, Zhang N, Feng J, Sun H-B. Graphene as a Transparent and Conductive Electrode for Organic Optoelectronic Devices. *Adv Electron Mater*. **2019**;5(10):1900247.

11. Ohta T, Bostwick A, Seyller T, Horn K, Rotenberg E. Controlling the Electronic Structure of Bilayer Graphene. *Science* (80). **2006**;313(5789):951–4.

12. McCann E. Asymmetry gap in the electronic band structure of bilayer graphene. *Phys Rev B*. **2006**;74(16):161403.

13. Min H, Sahu B, Banerjee SK, MacDonald AH. Ab initio theory of gate induced gaps in graphene bilayers. *Phys Rev B*. **2007**;75(15):155115.

14. Castro E V, Novoselov KS, Morozov S V, Peres NMR, dos Santos JMBL, Nilsson J, et al. Biased Bilayer Graphene: Semiconductor with a Gap Tunable by the Electric Field Effect. *Phys Rev Lett*. **2007**;99(21):216802.

15. Oostinga JB, Heersche HB, Liu X, Morpurgo AF, Vandersypen LMK. Gate-induced insulating state in bilayer graphene devices. *Nat Mater*. **2008**;7(2):151–7.

16. Zhang Y, Tang T-T, Girit C, Hao Z, Martin MC, Zettl A, et al. Direct observation of a widely tunable bandgap in bilayer graphene. *Nature*. **2009**;459(7248):820–3.

17. Mak KF, Lui CH, Shan J, Heinz TF. Observation of an Electric-Field-Induced Band Gap in Bilayer Graphene by Infrared Spectroscopy. *Phys Rev Lett*. **2009**;102(25):256405.

18. Oh JG, Pak K, Kim CS, Bong JH, Hwang WS, Im SG, et al. A High-Performance Top-Gated Graphene Field-Effect Transistor with Excellent Flexibility Enabled by an iCVD Copolymer Gate Dielectric. *Small*. **2018**;14(9):1–8.

19. Lemme MC, Member S, Echtermeyer TJ, Baus M, Kurz H. A Graphene Field-Effect Device. *IEEE Electron Device Letters*. **2007**;28(4):282–4.

20. Xu X, Wang C, Liu Y, Wang X, Gong N, Zhu Z, et al. A graphene P-N junction induced by single-gate control of dielectric structures. *J Mater Chem C*. **2019**;7(29):8796–802.

21. Chen FW, Ilatikhameneh H, Klimeck G, Chen Z, Rahman R. Configurable Electrostatically Doped High Performance Bilayer Graphene Tunnel FET. *IEEE J Electron Devices Soc*. **2016**;4(3):124–8.

22. Szafranek BN, Fiori G, Schall D, Neumaier D, Kurz H. Current Saturation and Voltage Gain in Bilayer Graphene Field Effect Transistors. *Nano Lett*. **2012**;12(3):1324–1328.

23. Xu K, Lu H, Kinder EW, Seabaugh A, Fullerton-Shirey SK. Monolayer Solid-State Electrolyte for Electric Double Layer Gating of Graphene Field-Effect Transistors. *ACS Nano*. **2017**;11(6):5453–64.

24. Xu K, Liang J, Woeppel A, Bostian ME, Ding H, Chao Z, et al. Electric Double-Layer Gating of Two-Dimensional Field-Effect Transistors Using a Single-Ion Conductor. *ACS Appl Mater Interfaces*. **2019**;11(39):35879–87.

25. Hayashi CK, Garmire DG, Yamauchi TJ, Torres CM, Ordonez RC. High On-Off Ratio Graphene Switch via Electrical Double Layer Gating. *IEEE Access*. **2020**;8:92314–21.

26. Efetov DK, Kim P. Controlling Electron-Phonon Interactions in Graphene at Ultrahigh Carrier Densities. *Phys Rev Lett*. **2010**;105(25):256805.

27. Campos R, Borme J, Guerreiro JR, Machado G, Cerqueira MF, Petrovykh DY, et al. Attomolar label-free detection of dna hybridization with electrolyte-gated graphene field-effect transistors. *ACS Sensors*. **2019**;4(2):286–93.

28. Son M, Kim H, Jang J, Kim SY, Ki HC, Lee BH, et al. Low-Power Complementary Logic Circuit Using Polymer-Electrolyte-Gated Graphene Switching Devices. *ACS Appl Mater Interfaces*. **2019**;11(50):47247–52.

29. Xiao J, Zhan H, Wang X, Xu ZQ, Xiong Z, Zhang K, et al. Electrolyte gating in graphene-based supercapacitors and its use for probing nanoconfined charging dynamics. *Nat Nanotechnol*. **2020**;15:683-689.

30. Lv R, Li Q, Botello-Méndez AR, Hayashi T, Wang B, Berkdemir A, et al. Nitrogen-doped graphene: Beyond single substitution and enhanced molecular sensing. *Sci Rep*. **2012**;2:1–8.

31. Wang H, Wang Q, Cheng Y, Li K, Yao Y, Zhang Q, et al. Doping monolayer graphene with single atom substitutions. *Nano Lett*. **2012**;12(1):141–4.

32. Rafique M, Shuai Y, Ahmed I, Shaikh R, Tunio MA. Tailoring electronic and optical parameters of bilayer graphene through boron and nitrogen atom co-substitution; an ab-initio study. *Appl Surf Sci*. **2019**;480:463–71.

33. Inani H, Mustonen K, Markevich A, Ding EX, Tripathi M, Hussain A, et al. Silicon Substitution in Nanotubes and Graphene via Intermittent Vacancies. *J Phys Chem C*. **2019**;123(20):13136–40.

34. Boas CRSV, Focassio B, Marinho E, Larrude DG, Salvadori MC, Leão CR, et al. Characterization of nitrogen doped grapheme bilayers synthesized by fast, low temperature microwave plasma-enhanced chemical vapour deposition. *Sci Rep*. **2019**;9(1):1–12.

35. Tadich A, Edmonds MT, Ley L, Fromm F, Smets Y, Mazej Z, et al. Tuning the charge carriers in epitaxial graphene on SiC(0001) from electron to hole via molecular doping with C60F48. *Appl Phys Lett*. **2013**;102(24):241601.

36. Szafranek BN, Schall D, Otto M, Neumaier D, Kurz H. High On/Off Ratios in Bilayer Graphene Field Effect Transistors Realized by Surface Dopants. *Nano Lett*. **2011**;11(7):2640–3.

37. Solís-Fernández P, Okada S, Sato T, Tsuji M, Ago H. Gate-Tunable Dirac Point of Molecular Doped Graphene. *ACS Nano*. **2016**;10(2):2930–9.

38. Iyakutti K, Kumar EM, Lakshmi I, Thapa R, Rajeswarapalanichamy R, Surya VJ, et al. Effect of surface doping on the band structure of graphene: a DFT study. *J Mater Sci Mater Electron*. **2016**;27(3):2728–40.

39. Zhou E, Xi J, Liu Y, Xu Z, Guo Y, Peng L, et al. Large-area potassium-doped highly conductive graphene films for electromagnetic interference shielding. *Nanoscale*. **2017**;9(47):18613–8.

40. Han T-H, Kwon S-J, Li N, Seo H-K, Xu W, Kim KS, et al. Versatile p-Type Chemical Doping to Achieve Ideal Flexible Graphene Electrodes. *Angew Chemie*. **2016**;128(21):6305–9.

41. Jørgensen AL, Duncan DA, Kastorp CFP, Kyhl L, Tang Z, Bruix A, et al. Chemically-resolved determination of hydrogenated graphene-substrate interaction. *Phys Chem Chem Phys*. **2019**;21(25):13462–6.

42. Ehlen N, Hell M, Marini G, Hasdeo EH, Saito R, Falke Y, et al. Origin of the Flat Band in Heavily Cs-Doped Graphene. *ACS Nano*. **2020**;14(1):1055–69.

43. Link S, Forti S, Stöhr A, Küster K, Rösner M, Hirschmeier D, et al. Introducing strong correlation effects into graphene by gadolinium intercalation. *Phys Rev B*. **2019**;100(12):121407.

44. Watcharinyanon S, I. Johansson L, Xia C, Ingo Flege J, Meyer A, Falta J, et al. Ytterbium Intercalation of Epitaxial Graphene Grown on Si-Face SiC. *Graphene.* **2013**;02(02):66–73.

45. McChesney JL, Bostwick A, Ohta T, Seyller T, Horn K, González J, et al. Extended van Hove Singularity and Superconducting Instability in Doped Graphene. *Phys Rev Lett*. **2010**;104(13):136803.

46. Ichinokura S, Sugawara K, Takayama A, Takahashi T, Hasegawa S. Superconducting Calcium-Intercalated Bilayer Graphene. *ACS Nano.* **2016**;10(2):2761–5.

47. Günther S, Menteş TO, Reichelt R, Miniussi E, Santos B, Baraldi A, et al. Au intercalation under epitaxial graphene on Ru(0001): The role of graphene edges. *Carbon*. **2020**;162:292–9.

48. Dedkov Y, Voloshina E. Spectroscopic and DFT studies of graphene intercalation systems on metals. *J Electron Spectros Relat Phenomena*. **2017**;219:77–85.

49. Daukiya L, Nair MN, Hajjar-Garreau S, Vonau F, Aubel D, Bubendorff JL, et al. Highly n-doped graphene generated through intercalated terbium atoms. *Phys Rev B*. **2018**;97(3):2–7.

50. Bonacum JP, O'Hara A, Bao DL, Ovchinnikov OS, Zhang YF, Gordeev G, et al. Atomic-resolution visualization and doping effects of complex structures in intercalated bilayer graphene. *Phys Rev Mater*. **2019**;3(6):1–10.

51. Rosenzweig P, Karakachian H, Link S, Küster K, Starke U. Tuning the doping level of graphene in the vicinity of the Van Hove singularity via ytterbium intercalation. *Phys Rev B*. **2019**;100(3):35445.

52. Braeuninger-Weimer P, Burton O, Weatherup RS, Wang R, Dudin P, Brennan B, et al. Reactive intercalation and oxidation at the buried graphene-germanium interface. *APL Mater*. **2019**;7(7).

53. Guo H, Zhang R, Li H, Wang X, Lu H, Qian K, et al. Sizable Band Gap in Epitaxial Bilayer Graphene Induced by Silicene Intercalation. *Nano Lett*. **2020**;20(4):2674–80.

54. Antoniazzi I, Chagas T, Matos MJS, Marçal LAB, Soares EA, Mazzoni MSC, et al. Oxygen intercalated graphene on SiC(0001): Multiphase SiOx layer formation and its influence on graphene electronic properties. *Carbon*. **2020**;167(15):746-759.

55. Briggs N, Bersch B, Wang Y, Jiang J, Koch RJ, Nayir N, et al. Atomically thin half-van der Waals metals enabled by confinement heteroepitaxy. *Nat Mater*. **2020**;19(6):637–43.

56. Wehenkel DJ, Bointon TH, Booth T, Bøggild P, Craciun MF, Russo S. Unforeseen high temperature and humidity stability of FeCl3 intercalated few layer graphene. *Sci Rep*. **2015**;5:1–5.

57. Kwon SJ, Han TH, Ko TY, Li N, Kim Y, Kim DJ, et al. Extremely stable graphene electrodes doped with macromolecular acid. *Nat Commun*. **2018**;9(1):1–9.

58. Piazza A, Giannazzo F, Buscarino G, Fisichella G, Magna A La, Roccaforte F, et al. Graphene p-Type Doping and Stability by Thermal Treatments in Molecular Oxygen Controlled Atmosphere. *J Phys Chem C*. **2015**;119(39):22718–23.

59. Kanahashi K, Tanaka N, Shoji Y, Maruyama M, Jeon I, Kawahara K, et al. Formation of environmentally stable hole-doped graphene films with instantaneous and high-density carrier doping via a boron-based oxidant. *npj 2D Mater Appl*. **2019**;3(1):1–7.

60. Sanders S, Cabrero-Vilatela A, Kidambi PR, Alexander-Webber JA, Weijtens C, Braeuninger-Weimer P, et al. Engineering high charge transfer n-doping of graphene electrodes and its application to organic electronics. *Nanoscale*. **2015**;7(30):13135–42.

61. Han KS, Kalode PY, Koo Lee Y-E, Kim H, Lee L, Sung MM. A non-destructive n-doping method for graphene with precise control of electronic properties via atomic layer deposition. *Nanoscale*. **2016**;8(9):5000–5.

62. Riedl C, Coletti C, Starke U. Structural and electronic properties of epitaxial graphene on SiC(0001): a review of growth, characterization, transfer doping and hydrogen intercalation. *J Phys D Appl Phys*. **2010**;43(37):374009.

63. Kotsakidis JC, Grubisic Cabo A, Yin Y, Tadich A, Myers-Ward R, DeJarld MT, et al. Freestanding n-doped Graphene via Intercalation of Calcium and Magnesium into the Buffer Layer - SiC(0001) Interface. *Chem. Mater.* **2020**;32(15):6464–6482.

64. Nyakiti LO, Wheeler VD, Garces NY, Myers-Ward RL, Eddy Jr. CR, Gaskill DK. Enabling graphene-based technologies: Toward wafer-scale production of epitaxial graphene. *MRS Bulletin*. **2012**;37(12).

65. Broekman L, Tadich A, Huwald E, Riley J, Leckey R, Seyller T, et al. First results from a second generation toroidal electron spectrometer. *J Electron Spectros Relat Phenomena*. **2005**;144–147:1001–4.

66. Mucha-Kruczyński M, Tsyplyatyev O, Grishin A, McCann E, Fal'ko VI, Bostwick A, et al. Characterization of graphene through anisotropy of constant-energy maps in angle-resolved photoemission. *Phys Rev B*. **2008**;77(19):195403.

67. Kresse G, Furthmüller J. Efficiency of ab-initio total energy calculations for metals and semiconductors using a plane-wave basis set. *Comput Mater Sci*. **1996**;6(1):15–50.

68. Perdew JP, Burke K, Ernzerhof M. Generalized Gradient Approximation Made Simple. *Phys Rev Lett*. **1996**;77(18):3865–8.

69. Grimme S, Ehrlich S, Goerigk L. Effect of the damping function in dispersion corrected density functional theory. *J Comput Chem*. **2011**;32(7):1456–65.

70. Chen M, Weinert M. Layer $k$-projection and unfolding electronic bands at interfaces. *Phys Rev B*. **2018**;98(24):245421.

71. Chen MX, Chen W, Zhang Z, Weinert M. Effects of magnetic dopants in Li0.8M0.2OH) FeSe (M=Fe, Mn, Co): Density functional theory study using a band unfolding technique. *Phys Rev B*. **2017**;96(24):245111.

72. Riedl C, Coletti C, Iwasaki T, Zakharov AA, Starke U. Quasi-free-standing epitaxial graphene on SiC obtained by hydrogen intercalation. *Phys Rev Lett*. **2009**;103(24):1–4.

73. Briggs N, Gebeyehu ZM, Vera A, Zhao T, Wang K, De La Fuente Duran A, et al. Epitaxial graphene/silicon carbide intercalation: A minireview on graphene modulation and unique 2D materials. *Nanoscale*. **2019**;11(33):15440–7.

74. Balog R, Cassidy A, Jørgensen J, Kyhl L, Andersen M, Čabo AG, et al. Hydrogen interaction with graphene on Ir(1 1 1): A combined intercalation and functionalization study. *J Phys Condens Matter*. **2019**;31(8).

75. Yurtsever A, Onoda J, Iimori T, Niki K, Miyamachi T, Abe M, et al. Effects of Pb Intercalation on the Structural and Electronic Properties of Epitaxial Graphene on SiC. *Small*. **2016**;(29):3956–66.

76. Mansour AE, Kirmani AR, Barlow S, Marder SR, Amassian A. Hybrid Doping of Few-Layer Graphene via a Combination of Intercalation and Surface Doping. *ACS Appl Mater Interfaces*. **2017**;9(23):20020–8.

77. Xu J, Dou Y, Wei Z, Ma J, Deng Y, Li Y, et al. Recent Progress in Graphite Intercalation Compounds for Rechargeable Metal (Li, Na, K, Al)-Ion Batteries. *Adv Sci*. **2017**;4(10):1700146.

78. Dresselhaus MS, Dresselhaus G. Intercalation compounds of graphite. *Adv Phys*. **2002**;51(1):1–186.

79. Emtsev K V., Zakharov AA, Coletti C, Forti S, Starke U. Ambipolar doping in quasifree epitaxial graphene on SiC(0001) controlled by Ge intercalation. *Phys Rev B*. **2011**;84(12):1–6.

80. Kim H, Dugerjav O, Lkhagvasuren A, Seo JM. Origin of ambipolar graphene doping induced by the ordered Ge film intercalated on SiC(0001). *Carbon*. **2016**;108:154–64.

81. Liu Y, Merinov B V, Goddard WA. Origin of low sodium capacity in graphite and generally weak substrate binding of Na and Mg among alkali and alkaline earth metals. *Proc Natl Acad Sci*. **2016**;113(14):3735–9.

82. Hüffner S. Photoelectron Spectroscopy. *Springer* 3rd ed. ISBN 978–3–662–09280–4.

83. Damascelli A, Hussain Z, Shen Z-X. Angle-resolved photoemission studies of the cuprate superconductors. *Rev Mod Phys*. **2003**;75(2):473–541.

84. Bostwick A, Ohta T, McChesney JL, Emtsev K V, Seyller T, Horn K, et al. Symmetry breaking in few layer graphene films. *New J Phys*. **2007**;9(10):385.

85. Shirley EL, Terminello LJ, Santoni A, Himpsel FJ. Brillouin-zone-selection effects in graphite photoelectron angular distributions. *Phys Rev B*. **1995**;51(19):13614–22.

86. Siegel DA, Park CH, Hwang C, Deslippe J, Fedorov A V., Louie SG, et al. Many-body interactions in quasi-freestanding graphene. *Proc Natl Acad Sci*. **2011**;108(28):11365–9.

87. Hwang C, Siegel DA, Mo SK, Regan W, Ismach A, Zhang Y, et al. Fermi velocity engineering in graphene by substrate modification. *Sci Rep*. **2012**;2:2–5.

88. McCann E, Koshino M. The electronic properties of bilayer graphene. *Reports Prog Phys*. **2013**;76(5):56503.

89. Taychatanapat T, Jarillo-Herrero P. Electronic Transport in Dual-Gated Bilayer Graphene at Large Displacement Fields. *Phys Rev Lett*. **2010**;105(16):166601.

90. Kanayama K, Nagashio K. Gap state analysis in electric-field-induced band gap for bilayer graphene. *Sci Rep*. **2015**;5(1):15789.

91. Gava P, Lazzeri M, Saitta AM, Mauri F. Ab initio study of gap opening and screening effects in gated bilayer graphene. *Phys Rev B*. **2009**;79(16):1–13.

92. Choi SM, Jhi SH, Son YW. Effects of strain on electronic properties of graphene. *Phys Rev B*. **2010**;81(8):23–6.

93. Klain C, Linde S, Shikler R, Sarusi G. Low work function Ca doped graphene as a transparent cathode for organic opto-electronics and OLEDs. *Carbon*. **2020**;157:255–61.

# Magnesium-intercalated graphene on SiC: highly n-doped air-stable bilayer graphene at extreme displacement fields

## Supplementary Information

*Antonija Grubišić Čabo[1†*], Jimmy C. Kotsakidis[1], Yuefeng Yin[2,3], Anton Tadich[3,4,5], Matthew Haldon[1], Sean Solari[1], Iolanda di Bernardo[1,3], Kevin M. Daniels[6], John Riley[5], Eric Huwald[5], Mark T. Edmonds[1,3], Rachael Myers-Ward[7], Nikhil V. Medhekar[2,3], D. Kurt Gaskill[7], Michael S. Fuhrer[1,3*]*

1. School of Physics and Astronomy, Monash University, Clayton VIC 3800, Australia

2. Department of Materials Science and Engineering, Monash University, Clayton VIC 3800, Australia

3. Centre for Future Low Energy Electronics Technologies, Monash University, Clayton VIC 3800, Australia

4. Australian Synchrotron, Clayton, VIC 3168, Australia

5. Department of Physics, La Trobe University, Bundoora VIC 3086, Australia

6. Department of ECE, University of Maryland, College Park, Maryland 20742, USA

7. U.S. Naval Research Laboratory, Washington D.C. 20375, USA

**Corresponding Author**

*Antonija@kth.se

* Michael.Fuhrer@monash.edu

**Present Address**

†KTH, Applied Physics, Stockholm SE 114 19, Sweden

## 1. Low-energy electron diffraction (LEED) of magnesium intercalated quasi-freestanding bilayer graphene (Mg-QFSBLG)

Figure S1 shows low-energy electron diffraction of quasi-freestanding bilayer graphene (Mg-QFSBLG) formed by intercalation of epitaxial monolayer graphene (EMLG) with magnesium, taken at an electron energy of 71 eV. Graphene coverage was nominally sub-monolayer. This data is taken on the same sample as shown in the main text, Figure 1-Figure 3. The Mg-QFSBLG LEED (Figure S1) shows additional ($\sqrt{3} \times \sqrt{3}$) R30° spots relative to the SiC lattice; these spots are not visible for the EMLG sample at the same electron energy, see Figure 1c in the main text. The ($\sqrt{3} \times \sqrt{3}$) R30° spots result from the interaction of magnesium with the SiC surface that causes reconstruction of the SiC. The ($\sqrt{3} \times \sqrt{3}$) R30° spots are more visible at higher electron energy that is more bulk sensitive, such as 126 eV, as in Figure 1d, compared to the more surface sensitive energy of 71 eV, as in Figure S1.

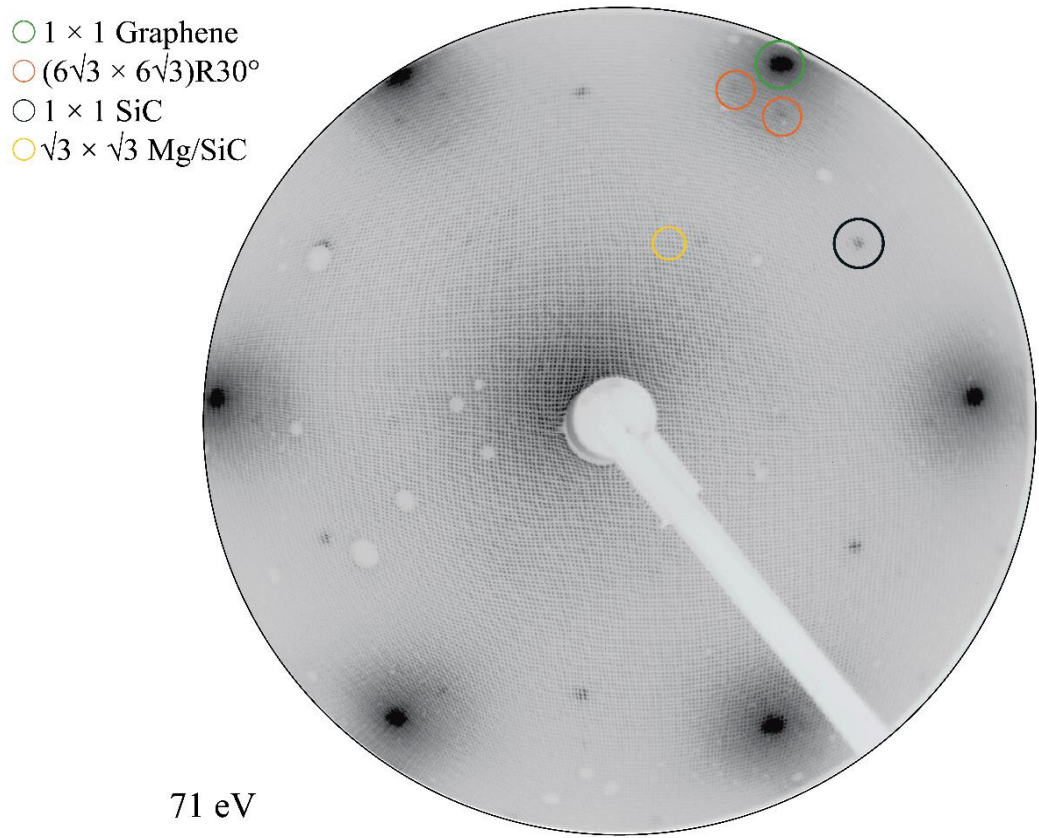

**Figure S1.** Low-energy electron diffraction of magnesium intercalated quasi-freestanding bilayer graphene (Mg-QFSBLG) taken at the 71 eV. (√3 × √3) R30° spots (yellow circle) arising from magnesium induced SiC reconstruction are less visible at this energy compared to more bulk sensitive energy of 126 eV. Green circle is graphene (1 × 1) spot, gray circle is SiC (1 × 1) spot and orange circles (6√3 × 6√3)R30° relative to the SiC are reconstruction arising due to the buffer layer.

## 2. Reduced momentum resolution in scans perpendicular to the $\bar{K} - \bar{\Gamma} - \bar{K}$ high-symmetry direction.

The toroidal analyser features a unique geometry which allows all polar emission angles ($\theta$ = -90° to +90°) to be measured along a high-symmetry azimuth ($\phi$). This is achieved via a concentric slit surrounding the sample. In the polar emission direction, measured along this given azimuth, the angular resolution of the detector is ~ 0.1°. However, the finite slit size at the sample position results in a small angular acceptance in the $\phi$ direction at each polar emission angle of ~ 1°. For this reason, data measured along the $\bar{K} - \bar{\Gamma} - \bar{K}$ high-symmetry direction, Figure S2, have smaller momentum broadening compared to the data measured along the direction perpendicular to the $\bar{K} - \bar{\Gamma} - \bar{K}$ high-symmetry direction, Figure 2 and Figure 4.

Tight-binding model for the bilayer graphene with an interlayer potential difference obtained by Pybinding [1] is overlaid on the data in Figure S2. This model used the same parameter as the low-energy approximate analytical solution in the main text: interlayer potential difference U = 0.87 eV, $\gamma_1$ = 0.4 eV, $\gamma_3$ = 0.12 eV, $a$ = 1.42 Å is the C-C distance in graphene, $a_{UC}$ = 2.46 Å is a unit cell length, $c$ = 3.35 Å is an interlayer separation, $v$ = 0.97 × 10$^6$ m/s is a band velocity and $\hbar$ is reduced Planck's constant.

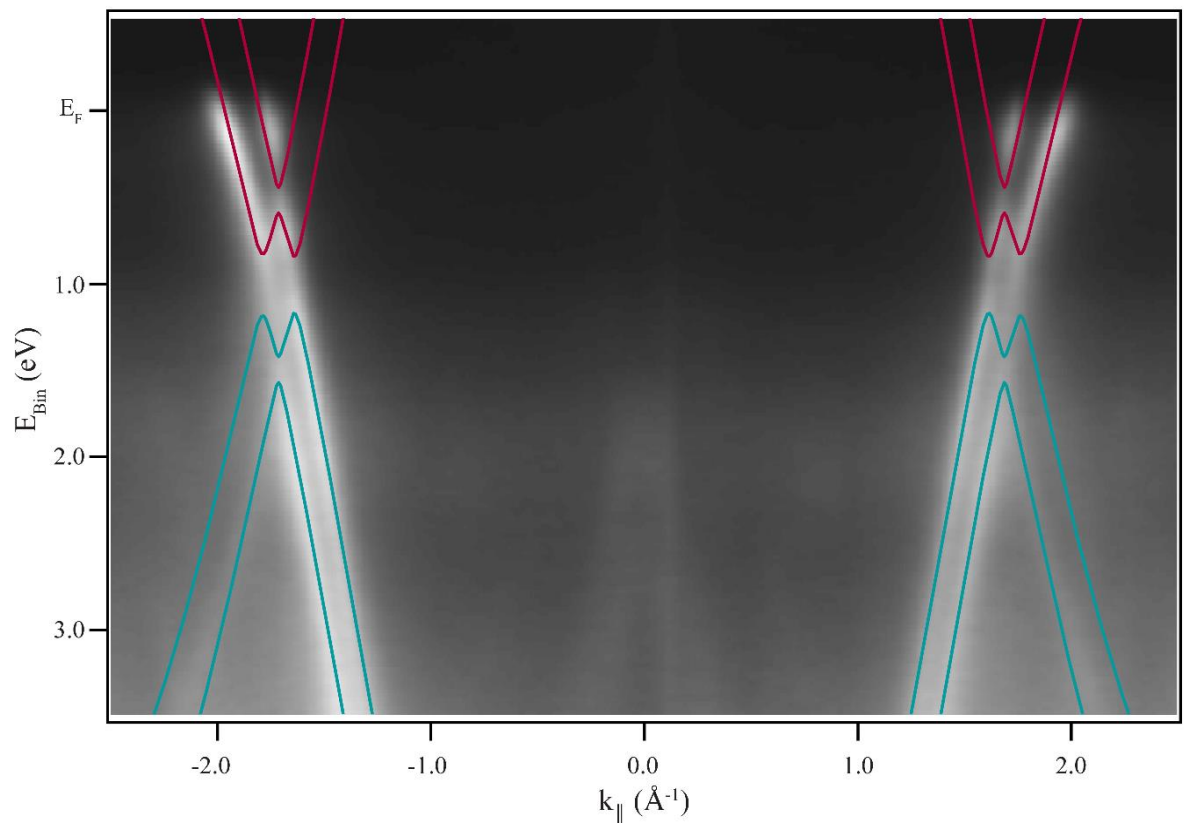

**Figure S2.** Electronic band structure along the $\bar{K} - \bar{\Gamma} - \bar{K}$ direction for magnesium intercalated sample. Overlaid bands in red and teal are from the tight-binding model for bilayer graphene with an interlayer potential difference of 0.87 eV.

## 3. Averaging angle-resolved photoemission spectroscopy (ARPES) data of Mg-QFSBLG taken along the K-G-K direction

When measuring graphene using linearly polarised light at normal incidence to the sample, an intensity suppression of half of the Dirac cone appears along the $\bar{K} - \bar{\Gamma} - \bar{K}$ direction due to a destructive interference between electrons coming from two sublattices of graphene [2]. In order to compare data taken along the $\bar{K} - \bar{\Gamma} - \bar{K}$ direction with the data taken along the direction perpendicular to the $\bar{K} - \bar{\Gamma} - \bar{K}$ direction, where both branches of the Dirac cone are visible (but which due to the mode of acquisition in which $k_{\parallel}$ is scanned using $\varphi$ instead of $\theta$, results in a lower momentum resolution), one needs both Dirac cone branches visible.

One way to obtain a full Dirac cone in this situation is by adding together the half cone data measured around the $\bar{K}$ and $\bar{K}'$ points; these can be summed together to produce a "full" Dirac cone, as shown in Figure S3. This is possible as although the $\bar{K}$ and $\bar{K}'$ points are inequivalent, they exhibit mirror symmetry in regards to the polarization effect and the intensity suppression is inverted for the two branches.

Note that this representation produces some artifacts; for example it exaggerates the effect of trigonal warping in the conduction bands (which are dominated by intensity from momenta along $\bar{K} - \bar{K}'$ direction) and suppresses the trigonal warping in the valence bands (which are dominated by intensity from momenta along $\bar{K} - \bar{\Gamma}$ direction).

Figure S4 shows ARPES data for Mg-QFSBLG, the same sample shown in Figure 1-3 of the main text, formed by intercalation of EMLG with magnesium, taken at the Toroidal Analyser Endstation, showing energy dispersion averaged along the $\bar{K} - \bar{\Gamma} - \bar{K}$. Dirac cones from inequevivalent $\bar{K}$ and $\bar{K}'$ points are added together to reconstruct a "full" Dirac cone, that looks very similar to the data obtained along the direction perpendicular to the $\bar{K} - \bar{\Gamma} - \bar{K}$ direction, including large splitting between inner and outer valence (conduction) band and the band gap

(see Figure 2 of the main text), but with much higher intensity and k-resolution. Data was taken with a photon energy of $h\nu$ = 100 eV and at room temperature.

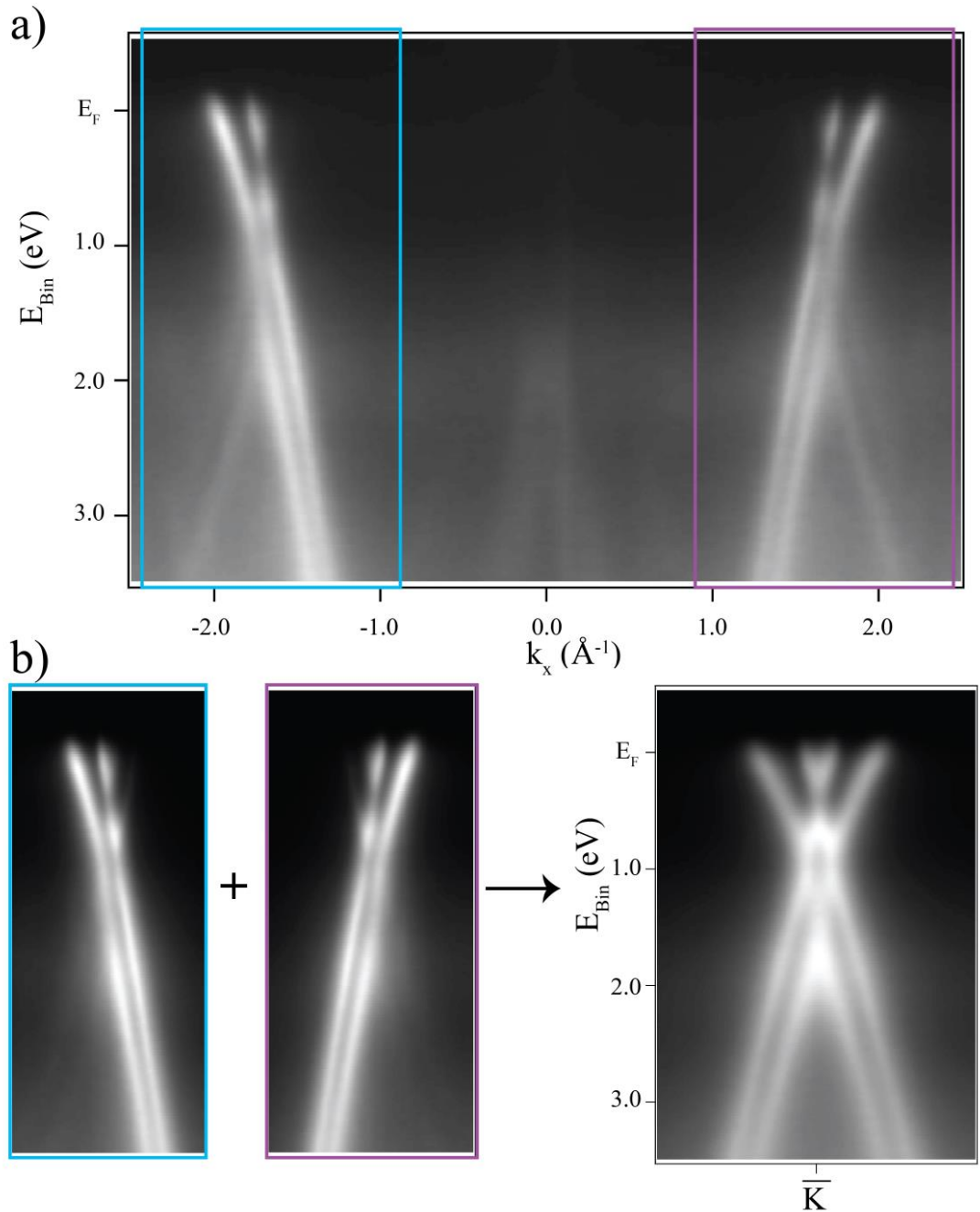

**Figure S3.** Angle-resolved photoemission spectroscopy data along the high symmetry direction $\bar{K} - \bar{\Gamma} - \bar{K}$ of Mg-QFSBLG is averaged to create a "full" Dirac cone. a) Only one side of a graphene Dirac cone is visible along $\bar{K} - \bar{\Gamma} - \bar{K}$. b) Data from $\bar{K}$ and $\bar{K}'$ point (left) is summed into a "full" Dirac cone (right).

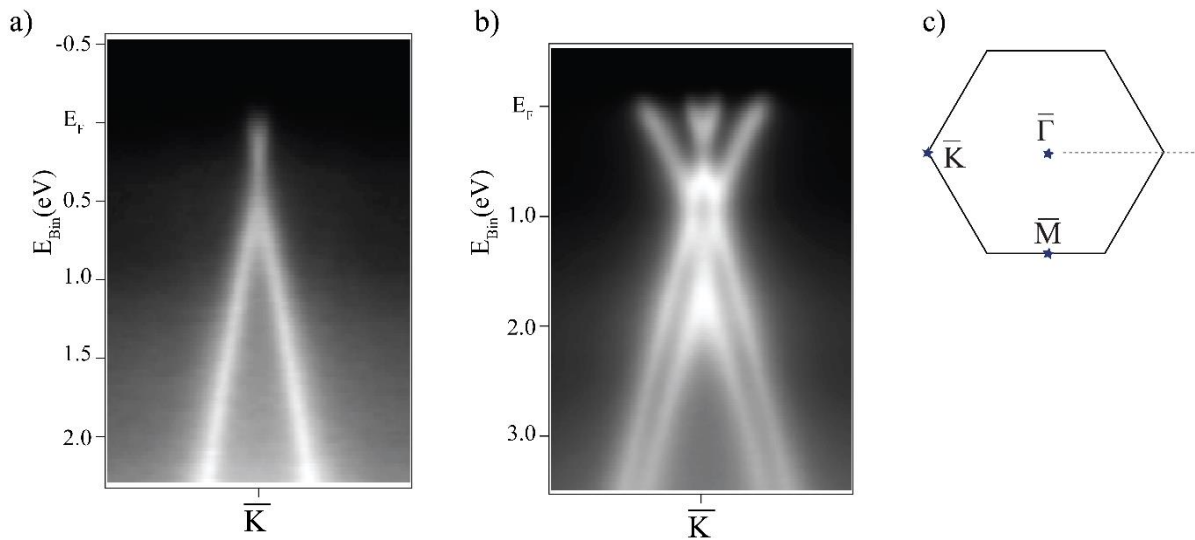

**Figure S4.** Angle-resolved photoemission spectroscopy data of a) EMLG and b) Mg-QFSBLG. Energy dispersion data is averaged along the $\overline{K} - \overline{\Gamma} - \overline{K}$ direction, shown in c).

### 3. LEED and ARPES of magnesium-intercalated quasi-freestanding graphene (2nd sample) upon exposure to air

An air exposure experiment was performed on a separate sample due to time constrains. The starting sample was nominally EMLG. Structural characterisation was done by LEED and electronic structure characterisation was done by ARPES, Figures S5-S7. No significant changes were observed in LEED and ARPES following air exposure.

After magnesium intercalation, EMLG is expected to be converted to Mg-QFSBLG. LEED data taken at 100 and 190 eV is shown in Figure S5. The (√3 x √3)R30° spots relative to SiC are visible, indicating magnesium reconstruction of the SiC surface and formation of Mg-QFSBLG.

Fermi surfaces and ARPES data along the direction perpendicular to the $\bar{K}-\bar{\Gamma}-\bar{K}$ direction is shown in Figure S6. Following intercalation, second set of bands appear in ARPES, however, it is less clear than in the case of first Mg-QFSBLG shown in the Figure 2. The tight binding (TB) model for bilayer graphene under high displacement field is overlaid in dark gray in Figure S6d and S6f. The same parameters were used as in the TB model in the main text ($v_F = 0.97 \times 10^6$ m/s, $U = 0.87$ V).

Data measured along the $\bar{K}-\bar{\Gamma}-\bar{K}$ high-symmetry direction for magnesium intercalated sample before and after air exposure is shown in Figure S7a and Figure S7b, respectively. Data is in agreement with the data along the direction perpendicular to the $\bar{K}-\bar{\Gamma}-\bar{K}$ high-symmetry direction.

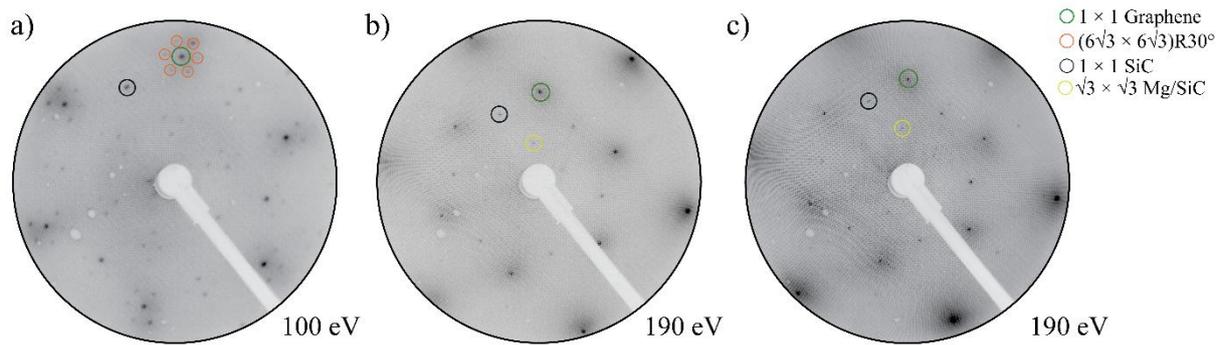

**Figure S5.** LEED characterization of second EMLG sample. Clean sample is shown in a), after magnesium intercalation in b) and after air exposure in c). Following magnesium intercalation, sample looks completely freestanding. No changes in LEED are observed following air exposure. Data taken at 100 eV (EMLG) and 190 eV (Mg-QFSBLG and air exposed Mg-QFSBLG) and room temperature. Green circle: (1 × 1) graphene lattice; gray circle: (1 × 1) SiC lattice; orange circles (6√3 × 6√3)R30° relative to SiC reconstructions arising from the buffer layer; yellow circle: (√3 ×√3) R30° reconstruction of SiC surface by magnesium.

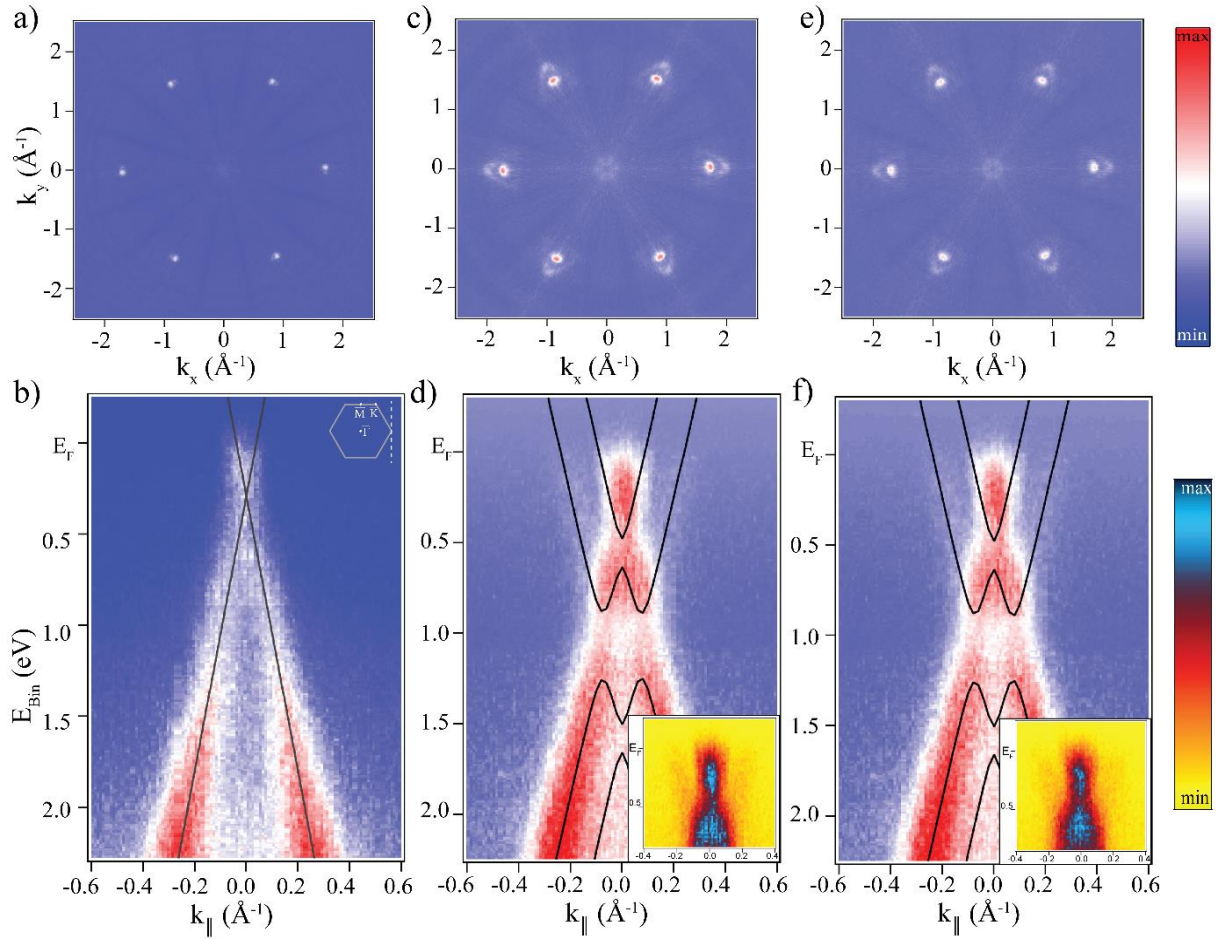

**Figure S6.** Angle-resolved photoemission spectroscopy data on the second Mg-QFSBLG graphene sample, formed by magnesium intercalation of an EMLG sample. Constant energy surface taken at the Fermi level and energy dispersion along the direction perpendicular to the $\bar{K} - \bar{\Gamma} - \bar{K}$ direction, shown in the inset in b), for the clean sample, a) and b), magnesium-intercalated sample, c) and d), and air exposed magnesium-intercalated sample, e) and f), respectively. Fermi surface data has been symmetrized around $\bar{K}$ point to increase signal-to-noise ratio. Insets in the Figure S6d and the Figure S6f show a region around the Fermi level where two sets of bands can be seen. Solid lines in b) are linear fit for the bandstructure of monolayer graphene, and solid lines in d), f) are a tight binding model of bilayer graphene using the same parameters as the main text.

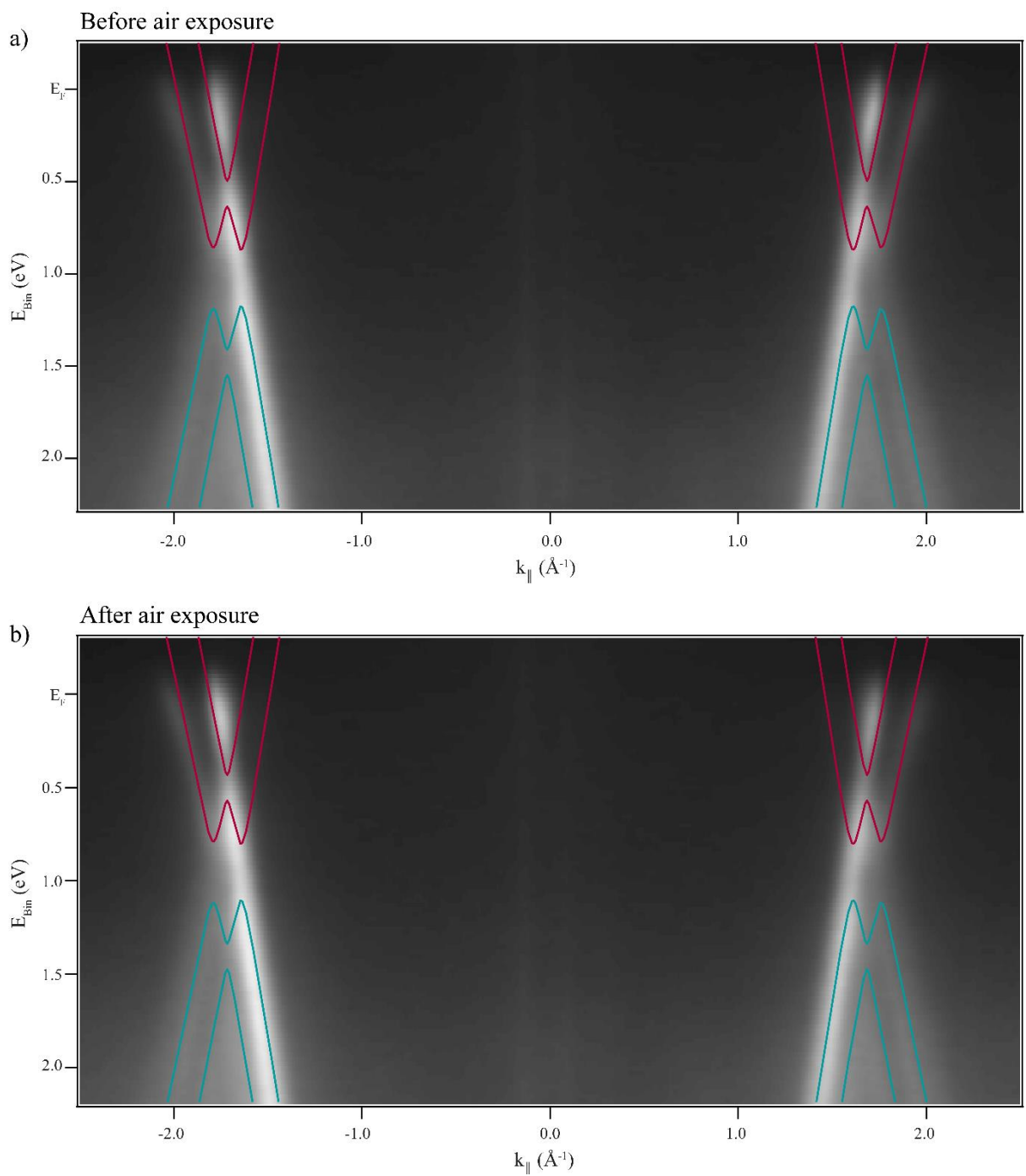

**Figure S7.** Electronic band structure along the $\bar{K} - \bar{\Gamma} - \bar{K}$ direction for magnesium intercalated sample before a) and after b) air exposure. Overlaid bands in red and teal are from the tight-binding model for bilayer graphene with an interlayer potential difference of 0.87 eV.

## 4. Additional ARPES data of Mg-QFSBLG (2$^{nd}$ sample) averaged along the K-G-K direction

Figure S8 shows additional ARPES data for the second Mg-QFSBLG, the same sample shown in Figure 4 in the main text, and in Figures S5-S7. Data is showing energy dispersion averaged along the $\overline{K} - \overline{\Gamma} - \overline{K}$ direction, with "full" Dirac cone visible. Following magnesium intercalation, Figure S8b, two sets of bands are visible indicating a bilayer formation. Compared to the first Mg-QFSBLG sample, shown in Figure S4b, where bands are of comparable intensity, in this sample the outer conduction band and inner valence band are less intense. Reason for this discrepancy is currently unknown. Data taken with a photon energy of $h\upsilon$ = 100 eV and at room temperature.

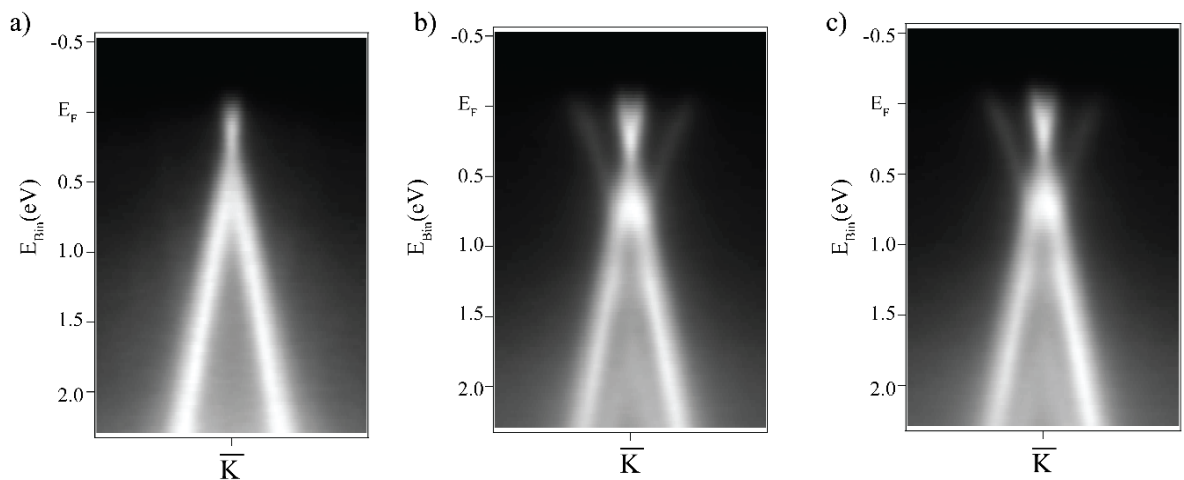

**Figure S8.** Angle-resolved photoemission spectroscopy data of a) pristine sample, b) sample after magnesium intercalation that formed Mg-QFSBLG and c) air exposed Mg-QFSBLG. Energy dispersion data is taken along the $\overline{K} - \overline{\Gamma} - \overline{K}$ direction.

## 5. First principles calculations of the energetics of Mg intercalation

The thermodynamic driving force for the conversion of EMLG on SiC to Mg-QFSBLG heterostructure shown in Figure 1b can be estimated by considering the reaction:

Monolayer graphene on SiC + Mg → Mg-QFSBLG on SiC.             (1)

Our first principles calculations on models shown in Figure 3a, 3b indicate that this reaction is energetically favourable by 1.18 eV. In our supercell models, the energy of Mg-QFSBLG on SiC and monolayer graphene on SiC is –291.44 eV and –288.98 eV, respectively, and we have used the energy of crystalline Mg (−1.28 eV per atom) as the reference. In each case, the supercell consisted of 3 layers of ($\sqrt{3} \times \sqrt{3}$) SiC crystal along with ($2 \times 2$) graphene layers.

We considered several initial adsorption sites of Mg on SiC, namely C-top, Si-top, C-Si-bridge and C-Si-center sites as shown in Figure S9. Mg adsorbed at C-top site is found to be stable and the most energetically favorable configuration. All other initial adsorption sites moved to C-top position upon structural relaxation (denoted by black arrows in Figure S9).

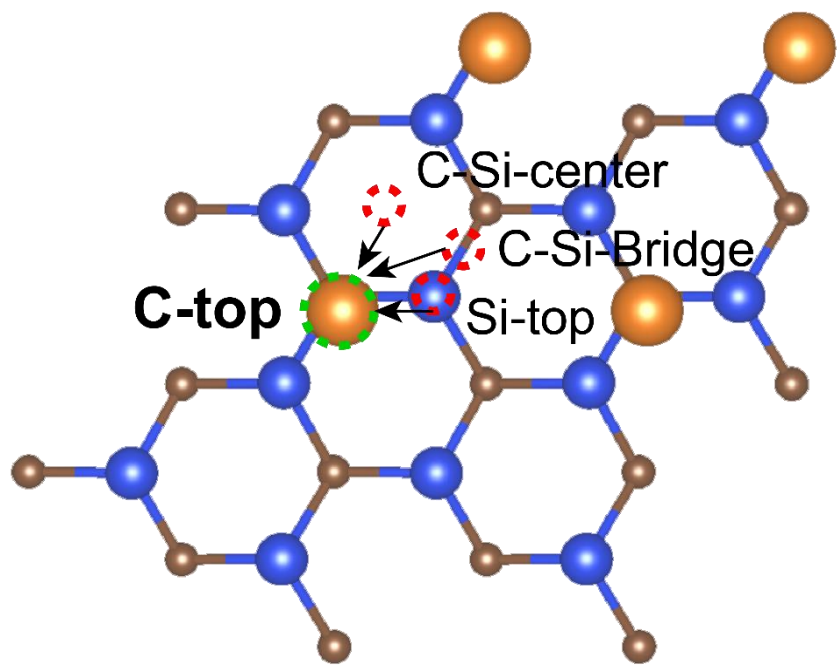

**Figure S9.** Top view of graphene/Mg/SiC interface. Brown, orange and blue spheres indicate carbon, magnesium and silicon atoms, respectively. The C-top site (green dashed circle) was found to be stable and energetically most favorable for magnesium adsorption. Other possible sites for magnesium shown by red dashed circles, namely, C-Si center, C-Si bridge, Si-top, were also considered.

References

(1)  Moldovan D, Anđelković M, Peeters F. pybinding v0.9.4: a Python package for tight-binding calculations. *Zenodo*; **2017**. Available from: https://doi.org/10.5281/zenodo.826942

(2)  Mucha-Kruczyński M, Tsyplyatyev O, Grishin A, McCann E, Fal'ko VI, Bostwick A, et al. Characterization of graphene through anisotropy of constant-energy maps in angle-resolved photoemission. *Phys Rev B*. **2008**;77(19):195403.